\documentclass{article}
\usepackage{hyperref}

\usepackage{graphicx}
\usepackage{amssymb}

\newcommand\eref[1]{(\ref{#1})}

\begin{document}

\title{Correlation properties of the random linear high-order Markov chains }

\author{V.~E.~Vekslerchik\footnote{\texttt{vekslerchik@yahoo.com}}, 
G.~M.~Pritula\footnote{\texttt{pritula.galina@gmail.com}}, 
S.~S.~Melnik\footnote{\texttt{melnik.teor@gmail.com}} 
and O.~V.~Usatenko}

\date{\it A. Ya. Usikov Institute for Radiophysics and
Electronics \\ Ukrainian Academy of Science, \\ 12 Proskura Street, 61805
Kharkiv, Ukraine}

\maketitle

\begin{abstract}
The aim of this paper is to study the correlation properties of
random sequences with additive linear conditional probability
distribution function (CPDF) and elaborate a reliable tool for their
generation. It is supposed that the state space of the sequence
under examination belongs to a finite set of real numbers. The
CPDF is assumed to be additive and linear with respect to the values
of the random variable. We derive the equations that relate the
correlation functions of the sequence to the memory function
coefficients, which determine the CPDF. The obtained analytical
solutions for the equations connecting the memory and correlation
functions are compared with the results of numerical simulation.
Examples of possible correlation scenarios in the high-order
additive linear chains are given.
\end{abstract}

\section{Introduction}
Our world is complex and correlated. The most peculiar
manifestations of this concept are human and animal communication,
written texts of natural languages, DNA nucleotide and protein
sequences, data flows in computer networks, stock indexes, solar
activity, weather, etc. For this reason, systems with long-range
interactions (and/or sequences with long-range memory) and natural
sequences with non-trivial information content have been the focus
of a large number of studies in different fields of science for the
past several decades.

Complexity of random sequences  is very often connected with
long-range correlations. This fact was demonstrated by studies in
many areas of contemporary
physics~\cite{bul,sok,bun,yan,maj,halvin},
biology~\cite{vossDNA,stan,buld,prov,yul,hao},
economics~\cite{stan,mant,zhang},
linguistics~\cite{schen,kant,kokol,ebeling,uyakm,uya,Mann}, chaotic
dynamical systems~\cite{Eren,Lind}, data compression~\cite{Salomon},
etc.

The studies of random systems in physical and engineering sciences
can be divided into two parts. The first one investigates, analyzes
and predicts the behavior of such systems, whereas the second one,
which is considerably smaller, develops the methods of generation of
random processes with desired statistical properties. This approach
provides not only a deeper insight into the nature of correlations
but is also a creative tool for designing the devices and appliances
with random components in their structure such as different
wave-filters, diffraction gratings, artificial materials, antennas,
converters, delay lines, etc. These devices can exhibit unusual
properties or anomalous dynamical, kinetic or transport
characteristics controlled by a proper choice of
disorder~\cite{IzrKrMak}.

There are many algorithms for generating long-range correlated
sequences: the Mandelbrot fast fractional Gaussian noise
generation~\cite{Mand71}, the Voss procedure of consequent random
addition~\cite{voss}, the correlated L\'evy walks~\cite{shl}, the
expansion-modification Li method~\cite{Li89}, the convolution method
~\cite{Makse}, the method of Markov
chains~\cite{RewUAMM,Anderson,Atayero}, etc.
If some
restrictions on possible states of random variables are
imposed, say, we need to generate a random dichotomous sequence,
then the problem becomes  more
complicated~\cite{uya,CarpStan1,CarpStan2,hod,nar1,nar2,IKMU,gener}.

In recent years, as a result of significant increase in computing
power and in connection with the problems of large data analysis,
Markov chains are literally
experiencing a burst of popularity in the most diverse fields of
science and technology. With the rise of the complexity of emerging problems, the simple Markov
property, when the conditional distribution of the subsequent state
of the chain depends only on the current state, becomes often
insufficient and the dependence of the subsequent state on the
previous $N$ states of the chain should be taken into account. Such
a generalization is referred to as a model of the Markov chain of
higher order, the $N$th order chain. For higher order chains,
obtaining exact analytical results or carrying out exhaustive
numerical calculations becomes practically impossible and one has to
resort to the construction of special models, such as, for example,
the model of the additive Markov chain~\cite{uya}.
In~\cite{uya,uyakm,RewUAMM} there was developed the model of linear
additive high-order dichotomic Markov chain that allows to generate
Markov sequences with prescribed statistical properties (given by
their 1st and 2nd moments) in an efficient way. The present paper
offers a generalization of this model to finite state Markov
sequences.

\section{High-order Markov chains}

Consider an infinite random stationary ergodic sequence
\begin{equation}\label{ranseq}
 \mathbb{S}= ..., X_{-1}, X_{0}, X_{1}, ... 
\end{equation}
where the random variables $X_{n}, \, n \in \mathbb{Z} $ take
values from a finite set $ \Omega$ of
real numbers.
We suppose that the random sequence $\mathbb{S}$ is a
\textit{high-order} \emph{Markov chain} ~\cite{Raftery,
Seifert,Shields}. The sequence $\mathbb{S}$ is the Markov chain if
it has the following property: the conditional probability
distribution function (CPDF) of random variable~$X_n$ to have a
certain value $x_n\in \Omega $ under the condition that \emph{all}
previous states are given depends only on $N$ previous states,
\begin{eqnarray}\label{def_mark}
 \mathbb{P}\left( X_{n}=x_{n} |..., X_{0}=x_{0},
 \, ... \, , X_{n-1}=x_{n-1} \right) \nonumber \\[8pt]
 =\mathbb{P}\left( X_{n}=x_{n} |X_{n-N}=x_{n-N}, \, ... \, , X_{n-1}=x_{n-1}
 \right).
\end{eqnarray}

Such sequences are also referred to as multi- or
$N$-step ones~\cite{uya,MUYaG,RewUAMM}. Sometimes the number $N$ is also
referred to as the \emph{order} or the \emph{memory length} of the
Markov chain. Hereafter we use the abbreviated notation:
\begin{displaymath}
 x_{n-N}, \, ... \, , x_{n-1}\equiv \left\{x_{i}\right\}_{n-N}^{n-1}.
\end{displaymath}

We consider the $N$-step Markov chain $ \mathbb{S}$ with a
\emph{linear} CPDF:
\begin{eqnarray}\label{def:lin0}
  &&\mathbb{P}\left( X_{n}=x_{n} |\left\{ X_{i}=x_{i}\right\}_{n-N}^{n-1}
				\right) 
		\nonumber \\[8pt]
  &&=  f(x_{n}; x_{n-N}, \, ... \,, x_{n-1})  \equiv
  f_{0}(x_{n}) + \sum_{m=1}^{N} f_{m}(x_{n}) x_{n-m}.
\end{eqnarray}
The additivity of the chain, presented here in the linear form,
means that the ``previous'' values $x_{n-N}, \, ... \,, x_{n-1}$
exert an independent linear effect on the probability of the value
of ``final``, generated variable $X_{n}=x_{n}$. 
The first term in the right-hand side of Eq.~(\ref{def:lin0}),
$f_{0}(x)$, is responsible for the correct reproduction of
statistical properties of uncorrelated sequences, the second one
containing weight functions $f_{m}(x), m=1,...,N $ takes into
account and correctly reproduces correlation properties of the chain
up to the second order. The higher order correlation functions
cannot be reproduced independently. We cannot control them and
reproduce correctly by means of the weight functions $f_{m}(x)$. If
all $f_{m}(x)$ are equal to zero, then the CPDF,
Eq.~\eref{def:lin0}, is nothing but  $f_{0}(x)$,  which becomes
the one-point probability function of the non-correlated chain.

There are some generic conditions which the CPDF should meet.
First, 
the value of the CPDF has to belong to the closed interval [0,1] for
any realization of the previous $N$ elements of the chain,
\begin{equation}\label{ergo_m}
 0 \leqslant \mathbb{P}\left( X_{n}=x_{n} |\left\{ X_{i}=x_{i}\right\}_{n-N}^{n-1}
\right)
 \leqslant 1, \, \, n \in \mathbb{Z}.
\end{equation}
We will return to this condition in Section \ref{Sec:Generation}.

Secondly, since the probability for the random variable $X_n$ to  take on any value from the state space $\Omega$  is equal to 1,  
the following equality should hold for all sets of variables:
\begin{eqnarray}
  \sum_{x_{n} \in\Omega} \mathbb{P}\left( X_{n}=x_{n} | \left\{ X_{i}=x_{i}\right\}_{n-N}^{n-1}
	\right)
    =  1,
\end{eqnarray}
which 
results in the corresponding restrictions on the functions $f_m(x), m=0,...,N$:
\begin{equation}
  \sum_{x_{n} \in\Omega} f_{0}(x_{n}) = 1,
  \hspace{10mm}
  \sum_{x_{n} \in\Omega} f_{m}(x_{n}) = 0,
  \hspace{5mm}
  m = 1, \, ... \,, N.
    \label{sum:f}
\end{equation}
The right-hand side of Eq.~\eref{def:lin0} can be considered as  two first terms of
expansion of the function $\mathbb{P}\left( .| .\right) $ in a
series with respect to the ``small'' values $x_n$.

\section{One-point probability distribution
function
and averages}

The CPDF determines all statistical properties of a random
sequence. In this section, for the sequence given by Eq.~\eref{def:lin0},
we obtain the simplest and most common statistical characteristics -- 
the one-point probability distribution function and the average value of random variable, which then will be used  in Sections \ref{CorrEq} and \ref{sec:BC} for
deriving the equations for the correlation functions of the sequence.
The one-point probability  $\mathbb{P}(X_{n}=x_{n})$  can be expressed in terms
of the weight function $f_m(x)$ as follows:
\begin{eqnarray}
 && \mathbb{P}( X_{n}=x_{n} ) =
  \nonumber \\[8pt]
  && =
  \sum_{x_{n-N}, \, ... \,, x_{n-1} \in\Omega}
  \mathbb{P}\left( \left\{ X_{i}=x_{i}\right\}_{n-N}^{n-1} \right)
  \mathbb{P}\left( X_{n}=x_{n} | \left\{ X_{i}=x_{i}\right\}_{n-N}^{n-1}
	\right)\nonumber \\[8pt]
  && =
  \sum_{x_{n-N}, \, ... \,, x_{n-1} \in \Omega}
  \mathbb{P}\left( \left\{ X_{i}=x_{i}\right\}_{n-N}^{n-1}  \right)
  \biggl[ f_{0}(x_{n}) + \sum_{m=1}^{N} f_{m}(x_{n}) \, x_{n-m} \biggr]\nonumber \\[8pt]
\nonumber \\[8pt]
  && =  f_{0}(x_{n})  +  \sum_{m=1}^{N} f_{m}(x_{n})\sum_{x_{n-m} \in \Omega}
    \mathbb{P}\left( X_{n-m}=x_{n-m}\right)x_{n-m},
\end{eqnarray}
where $\mathbb{P}\left( \left\{ X_{i}=x_{i}\right\}_{n-N}^{n-1} \right)$ is the
joint probability function. Thus,
\begin{equation}\label{point-prob}
  \mathbb{P}( X_{n}=x_{n} )  =
  f_{0}(x_{n}) + \langle X \rangle \sum_{m=1}^{N} f_{m}(x_{n}),
\end{equation}
and
\begin{equation}\label{point-prob1}
 \langle X \rangle= \langle X_n \rangle= \sum_{x_{n} \in \Omega}
  \mathbb{P}( X_n=x_{n} ) \, x_{n}
\end{equation}
determines the average value of the random variable. 
The angles $\langle ... \rangle$ mean a statistical average over an
ensemble of sequences. When it comes to the numerical construction
of random chain, this average will be replaced by the average along the chain
due to ergodicity of the sequence. Due to stationarity of the
sequences under consideration, $\langle X \rangle$ is
$n$-independent. To obtain a solution of Eqs.~\eref{point-prob} and
\eref{point-prob1}, we multiply Eq.~\eref{point-prob} by $x$ and
average it (take the sum over $x$),
\begin{equation}
  \sum_{x \in \Omega} \mathbb{P}( X=x ) \, x
  =
  \sum_{x \in \Omega}f_{0}(x) x + \langle X \rangle \sum_{n=1}^{N}\sum_{x \in \Omega}
  f_{n}(x)x,
\label{point-prob2}
\end{equation}
from where we find 
\begin{equation}
  \langle X\rangle
  =\frac{\sum_{x \in \Omega}f_{0}(x) x }{1-\sum_{n=1}^{N}\sum_{x \in \Omega}
  f_{n}(x)x}.
\label{point-prob3}
\end{equation}
Definitely, we have $\mathbb{P}( X=x )$,
\begin{eqnarray}\label{onePP}
  \mathbb{P}( X=x )
  =
  f_{0}(x)
    +
    \frac{\sum_{x' \in \Omega}f_{0}(x') x' }{1-\sum_{m=1}^{N}\sum_{x' \in \Omega} f_{m}(x')x'}
   \sum_{m=1}^{N} f_{m}(x),
\end{eqnarray}
expressed in terms of $f_{m}(x), \, m=0, 1,..., N$, only. It follows
from Eq.~\eref{onePP}

\textbf{Proposition 1}. If a distribution function $\mathbb{P}( X=x
)$ of non-correlated sequence is $ f_{0}(x)$ with zero mean value,
$\langle X\rangle=\sum_{x \in \Omega} f_{0}(x)x =0$, then the
additional terms in Eq.~\eref{def:lin0} proportional to $f_m(x), \;
m=1,...,N,$ and describing correlations in the chain, do not change
the one-point distribution function.

In some cases it is convenient to write down Eq.~\eref{def:lin0} in
the form,
\begin{eqnarray}\label{def_lin1}
  \mathbb{P}\left( X_{n}=x_{n} |\left\{ X_{i}=x_{i}\right\}_{n-N}^{n-1}  \right)  =
  f_{0}(x_n) + \sum_{m=1}^{N} f_{m}(x_n) \left( x_{n-m}- \langle
  X\rangle\right),
\end{eqnarray}
providing the above formulated property because of $\langle
x_{n}- \langle
  X\rangle \rangle=0$.

\section{Recurrence equations for the pair correlators}\label{CorrEq}

In this section we derive the equation for the correlation functions
\begin{equation}\label{C(n)}
  C(n) = K(n) - \langle X\rangle ^{2},
\end{equation}
where $K(n)$ is
\begin{equation}\label{K(n)}
  K(n)  = \langle x_{0} x_{n}\rangle = \sum_{x_{0}, \, ... \,, x_{n} \in \Omega}  x_{0} \,
  \mathbb{P}\left(\left\{ X_{i}=x_{i}\right\}_{0}^{n} \right)  x_{n}.
\end{equation}

In view of the symmetry of the correlation function $C(-n)= C(n)$,
we can restrict our consideration with the positive values of $n$. The fact
that the correlator does not depend on the positions $n$ and  $m$ of
random numbers $x_{n}$ and $x_{m}$, but only on the distance between
them, $C(n,m)= C(n-m)$, is a consequence of stationarity of the
chain.

\textbf{Proposition 2}. For all $n \geqslant N$ the correlation functions $
C(n)$ satisfy the recurrence equations,
\begin{equation}\label{eq:corrC}
  C(n) = \sum_{m=1}^{N} F_{m} C(n-m),
\end{equation}
with the memory functions $F_{m}$ determined by the weight
functions $f_{m}(x)$,
\begin{equation}
  F_{m} = \sum_{x \in \Omega} x f_{m}(x),
  \qquad
  m = 1, \, ... \,, N.
\end{equation}
To prove this, let us consider the expression for $K(n)$. By
definition \eref{K(n)} we have
\begin{eqnarray}
& K(n) &  =  \sum_{x_{0}, \, ... \,, x_{n} \in \Omega}  x_{0} \,
  \mathbb{P}\left(\left\{ X_{i}=x_{i}\right\}_{0}^{n} \right)  x_{n}\nonumber\\[8pt]
 & =& \!\!\!\!\!\!\!\! \sum_{x_{0}, \, ... \,, x_{n-1} \in \Omega}  x_{0} \,
  \mathbb{P}\left(\left\{ X_{i}=x_{i}\right\}_{0}^{n-1} \right) 
  \sum_{ x_{n} \in \Omega}   x_{n}
  \left[    f_{0}( x_{n})    +    \sum_{m=1}^{N} f_{m}(x_{n}) x_{n-m}  \right]\nonumber\\[8pt]
  & = &\!\!\!\!\!\!\!\!  \sum_{x_{0}, \, ... \,, x_{n-1} \in \Omega}  x_{0} \,
  \mathbb{P}\left(\left\{ X_{i}=x_{i}\right\}_{0}^{n-1} \right)
    \left[    F_{0}    +    \sum_{m=1}^{N} F_{m} x_{n-m}  \right],
\label{corr-b}
\end{eqnarray}
where
\begin{equation}
  F_{\mu} = \sum_{x \in \Omega} x f_{\mu}(x),
  \qquad
  \mu = 0, \, ... \,, N.
    \label{def:F}
\end{equation}
In terms of $F$, expression \eref{corr-b} reads
\begin{equation}
  K(n) =  F_{0} \langle X \rangle + \sum_{m=1}^{N} F_{m} K(n-m).
\label{corr-c}
\end{equation}
Using Eqs.~\eref{point-prob} and \eref{point-prob1}, one can get $\langle X \rangle = F_{0}
+ \langle X \rangle \sum_{m=1}^{N} F_{m}$, or
\begin{equation}
  F_{0}  =  \langle X \rangle \left[ 1 - \sum_{m=1}^{N} F_{m} \right]\nonumber
\end{equation}
which converts \eref{corr-c} into
\begin{equation}
  K(n) =
  \langle X\rangle ^{2} \left[ 1 - \sum_{m=1}^{N} F_{m} \right]
  +  \sum_{m=1}^{N} F_{m} K(n-m).
\end{equation}
It is easy to see that rewriting the last equation in terms of
$C(n)$, $C(n) = K(n) - \langle X \rangle ^{2}$, one can obtain
the nice equation in the form of Eq.~\eref{eq:corrC}.

These recurrent equations determine the value of the correlation
function $C(n)$ at $n \geqslant N$ by its $N$ previous values $C(n-m),
\,\, m=1,...,N$.

\section{Boundary equations \label{sec:BC}}

During the derivation of the recurrence Eqs.~\eref{eq:corrC} we nowhere used the
condition $n \geqslant N$. This implies they are valid for $n < N$ as well,
but their meaning is completely different. Now Eqs.~\eref{eq:corrC}
connect different correlation functions $C(n)$ with negative and
positive arguments $n$:

\textbf{Proposition 3}. For $|n| \leqslant N-1$
equations~\eref{eq:corrC} in the form
\begin{equation}\label{eq:corrC1}
  C(n) = \sum_{m=1}^{N} F_{m} C(|n-m|)
\end{equation}
are the boundary conditions. Their solution determines the
correlation functions $C(1),...,C(N-1)$  as functions of $C(0)$ and
$F_{1}, ...,F_{N}$, where $C(0)=\langle(X-\langle X\rangle
)^2\rangle$ is the average with respect to the one-point probability function
$\mathbb{P}( X=x )$, Eq.~\eref{onePP}.

For $|n| \leqslant N-1$, with account of the symmetric property of
correlation function, Eqs.~\eref{eq:corrC1} represent a closed
system of $N-1$ inhomogeneous difference equations with constant
coefficients:
\begin{equation}\label{corrcycled0}
    \textbf{H}\left(\begin{array}{c} C(1) \\ \vdots \\ C(N-1) \end{array}\right)
  = - C(0)\left(\begin{array}{c} F_{N-1} \\ \vdots \\ F_{1} \end{array}\right),
\end{equation}
where the matrix $\textbf{H}$ can be represented as the sum of the
Hankel matrix $\textbf{F}^{(1)}$ and the Toeplitz matrix
$\textbf{F}^{(2)}$ ,
\begin{equation}\label{corrcycled1}
 \textbf{F}_{ij}^{(1)}=
    \left\{
    \begin{array}{ll}
        F_{N-i-j}, & i+j<N,\\
    -1, & i+j=N,\\
    0, & i+j>N,
    \end{array}
        \right.
\end{equation}
and
\begin{equation}\label{corrcycled2}
 \textbf{F}_{ij}^{(2)}=
    \left\{
    \begin{array}{ll}
        F_{N-i+j}, & i\geqslant j,\\
    0, & i<j.
    \end{array}
        \right.
\end{equation}

The solutions of the system \eref{corrcycled0} have the form
\begin{equation}
  C(n) =
        \Gamma_{n}(F_1,...,F_N) C(0), \quad n=1,...,N-1,
\label{bc:corrC}
\end{equation}
with uniquely defined constants $\Gamma_{n}(F_1,...,F_N)$:
\begin{equation}
    \Gamma_{n}(F_1,...,F_N)=
            - \sum_{k=1}^N{(H^{-1})_{n,k} F_{N-k}}.
\label{bc:corrCC}
\end{equation}
Thus, the problem of determining the pair correlation functions of
the $N$th order additive Markov chain with a given weight
coefficients $f_{m}(x), \;  m = 1, \, ... \,, N$, is reduced to
the Cauchy problem for the  $N$-order difference equation
\eref{eq:corrC} with $N$ border conditions \eref{bc:corrC}.

\section{General solution of the equation for the correlators}

Equation~\eref{eq:corrC} for the correlation function for $n \geqslant N$,
is a linear difference equation with constant coefficients (see, for
example, \cite{Elaydi}). It has particular solutions,
\begin{equation}
  C(n,\lambda) = \lambda^{n},
\end{equation}
provided $\lambda$'s are solutions of the polynomial equation
\begin{equation}
  \lambda^{N} = \sum_{m=1}^{N} F_{m} \lambda^{N-m}.
    \label{CharEq}
\end{equation}
Since an $N$-order polynomial has $N$ roots, $\lambda_{j}, j=1,...,N$,
the general solution of Eq.~\eref{eq:corrC} is given
by the linear combination of the corresponding particular solutions,
\begin{equation}
  C(n) = \sum_{j=1}^{N} \gamma_{j} \lambda_{j}^{n}.
\label{GenSol}
\end{equation}

The constants $\gamma_{j}$ should be determined from the boundary
conditions (see Section \ref{sec:BC}). To this end, we rewrite the general solution
\eref{GenSol} for $n=0,...,N-1$ in the matrix form,
\begin{equation}\label{V}
  \textbf{V}
\left(\begin{array}{c} \gamma_{1} \\ \vdots \\ \gamma_{N}
\end{array}\right)
  =
  \left(\begin{array}{c} C(0) \\ \vdots \\ C(N-1) \end{array}\right),
\end{equation}
where $\textbf{V}$ is the Vandermonde matrix,
\begin{equation}\label{VdM}
  \textbf{V}=
    \left(
  \begin{array}{ccc}
    1                 & \dots  & 1 \\
    \vdots            & \ddots & \vdots \\
   \lambda_{1}^{N-1} & \dots  & \lambda_{N}^{N-1}
    \end{array}
    \right),
\end{equation}
and recall Eq.~\eref{bc:corrC} which expresses $C(1),...,C(N-1)$ in terms of
$C(0)$. This leads to the following result:
\begin{equation}\label{gamma}
    \gamma_j=
            C(0) \sum_{k=1}^N{(\textbf{V}^{-1})_{jk} \Gamma_{k-1}}, \quad j=1,...,N,
\end{equation}
 where $\Gamma_j$ is determined in \eref{bc:corrCC}.

The natural requirement of vanishing of correlations as
$n\rightarrow\infty$ leads to constraints for the coefficients
$F_m$.  It is clear from solution \eref{GenSol} that, in order to
meet this requirement, all roots of the characteristic polynomial
should be located on the complex plane inside the unite circle,
$|{\lambda}|=1$ .

The problem of the polynomial roots distribution with respect to the
unit circle often arises in many applied problems, for example, the
ones of automatic control, digital signal processing, system
identification etc. There are various methods and algorithms for
solving this problem, the most widely used of which are the Schur-Cohn,
Juri, Bistritz procedures and their various modifications, see, for
example, \cite{Stoica,Bistritz}.

Summarizing the consideration of the two previous Sections, we can
formulate

\textbf{Proposition 4}. Statistical properties of the chain of
random variables $X_{n}$ generated by means of the CPDF defined by
Eq.~\eref{def:lin0} are described by the pair correlation functions:
\begin{equation}\label{RezCorr}
 C(n)=
    \left\{
    \begin{array}{ll}
        -C(0)\sum_{k=1}^N{(H^{-1})_{nk} F_{N-k}}, & n \leqslant N -
        1,\\ [16pt]
   -C(0)\sum_{j,k,m=1}^{N} {(\textbf{V}^{-1})_{jk}
    {(H^{-1})_{(k-1)m} F_{N-m}}
    } \lambda_{j}^{n}, & n\geqslant N.
    \end{array}
        \right.
\end{equation}

In the next section we analyze the solutions \eref{RezCorr} for the chains 
of the lowest orders and give examples of possible correlation functions
for higher order chains.

\section{Correlation functions of the chain with a given memory
function}

\subsection{The lowest order chains}
For \textit{the 1st order chain}, \eref{eq:corrC} takes the form
$C(n)=F_1C(n-1)$. It is easily solved  and its solution, which
satisfies the symmetry condition for the correlator, is the function
$C(n)=C(0) {F_1}^{|n|}$ exponentially decreasing when $|F_1|< 1$.

\begin{figure}
\begin{center}
  \includegraphics[width=55mm]{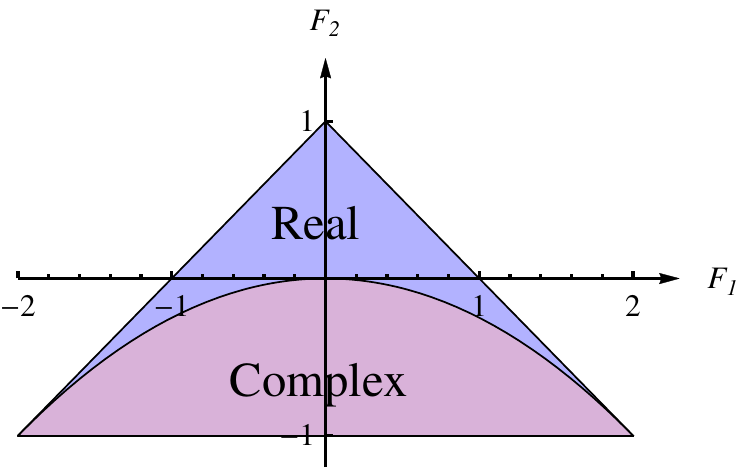}
\end{center}
  \caption{The range of the memory functions for which the
condition  $C(n)\rightarrow\ 0$ as $n\rightarrow\infty$ is satisfied
(when the zeros of the characteristic polynomial are inside the circle 
$|\lambda|<1$ on the complex
plane).}
  \label{fig:1}
\end{figure}

In the case of \textit{the 2nd order chain}, it is easy to show that the
necessary and sufficient condition for the roots of the quadratic
characteristic polynomial $p_2=\lambda^2-F_1 \lambda - F_2 $ to be
inside the unit circle on the complex plane is $|F_1|<1-F_2<2$, i.e.
the coefficients of the polynomial take values from the region
delineated by the triangle depicted in Fig. \ref{fig:1}. When
$F_2\geqslant -F_1^2/4$ , the roots are real, otherwise two of them are
complex and conjugate and third is real.

In the case of the 2nd order chain, system of equations \eref{corrcycled0} reduces to a single equation
\begin{equation}
    C(1)=F_1 C(0)+F_2 C(0),
 \end{equation}
and, therefore, there is just one constant $\Gamma_1$ (see \eref{bc:corrC}):
\begin{equation}
   \Gamma_{1}(F_1,F_2)=F_1(1-F_2)^{-1}.
 \end{equation}

\begin{figure}
\begin{center}
  \includegraphics[width=120mm]{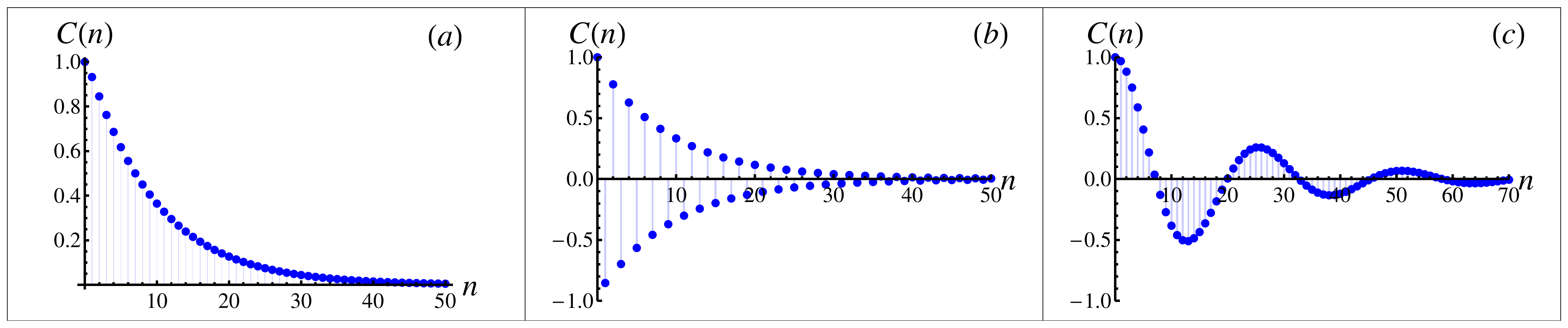}
\end{center}
  \caption{The correlation function $C(n)$ of the 2nd order sequence
with different memory functions: a) $F_1=1,1$ and
 $F_2= -0,18$; b) $F_1= -0,7$ and $F_2= 0,18$; c) $F_1=1,84$ and $F_2=-0,9$.}
  \label{fig:2}
\end{figure}

Solution \eref{GenSol} for the 2nd order chain
is
\begin{equation}
    C(n)=\gamma_{1}{\lambda_1}^n+\gamma_{2}{\lambda_2}^n,
    \hspace{10mm}
    \lambda_{1,2}=\frac{F_1\pm \sqrt{F_1^2+4F_2}}{2},
\end{equation}
where $\gamma_{1,2}$ are
\begin{equation}
  \gamma_1=C(0) \frac{\lambda_2-\Gamma_1}{\lambda_2-\lambda_1},
  \hspace{10mm}
  \gamma_2=C(0) \frac{\Gamma_1-\lambda_1}{\lambda_2-\lambda_1}.
\end{equation}
Correlation functions for different values of memory functions $F_1$
and $F_2$ are presented in Fig. \ref{fig:2}.

\begin{figure}
\begin{center}
  \includegraphics[width=60mm]{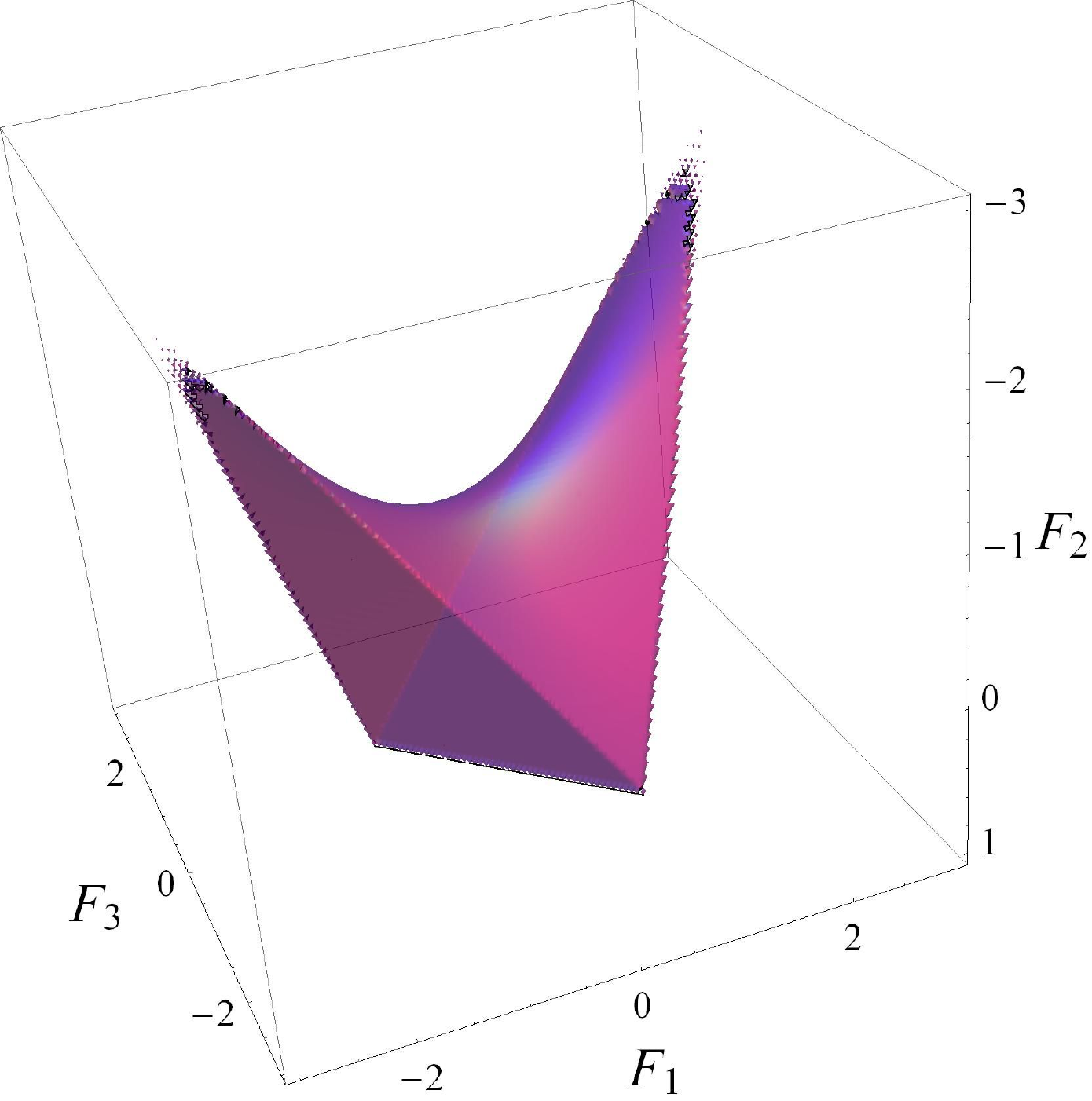}
\end{center}
  \caption{The range of the memory functions for which the
condition  $C_n\rightarrow\ 0$ as $n\rightarrow\infty$ is satisfied
(when the zeros of the characteristic polynomial $p_3$ are inside 
the circle $|\lambda|<1$ on the complex
plane).}
  \label{fig:3}
\end{figure}

In the case of the \textit{3rd order chain}, for
the roots of the cubic characteristic polynomial $p_3=\lambda^3-F_1
\lambda^2 - F_2 \lambda -F_3$ to lie inside the unit circle on the
complex plane, it is necessary and sufficient that the following
conditions be satisfied \cite{Grove}:
\begin{equation}
  |F_1+F_3|<1-F_2,
  \hspace{10mm}
  |F_1-3 F_3|<3+F_2,
  \hspace{5mm}
{F_3}^2-F_2-F_1 F_3<1.
\end{equation}
The corresponding range of allowed values of the memory function
is shown in Fig. \ref{fig:3}.

For the 3rd order chain the system \eref{corrcycled0}, determining the
boundary conditions,  has a form
\begin{equation}
  \left(
  \begin{array}{cc}
    - F_{1} - F_{3} & 1\\
    1 - F_{2}       & - F_{3}
       \end{array}
  \right)
  \left( \begin{array}{c}
    C(1) \\ C(2)
    \end{array}
    \right)
  =
  C(0)
  \left( \begin{array}{c} F_{2} \\ F_{1} \end{array} \right)
\end{equation}
and constants $\Gamma_{1,2}$, \eref{bc:corrCC}, are
\begin{equation}
  \Gamma_1(F_1,F_2,F_3)= \frac{F_{1}+F_{2}F_{3}}{\Delta},
  \hspace{10mm}
  \Gamma_2(F_1,F_2,F_3)= \frac{F_{1}^2+F_{1}F_{3}-F_{2}^2}{\Delta},
\end{equation}
where
\begin{equation}
  \Delta=1-F_{2}-(F_1+F_3)F_3.
\end{equation}
 Solution \eref{GenSol} for the 3rd order chain is
\begin{equation}
    C(n)=\gamma_{1}{\lambda_1}^n+\gamma_{2}{\lambda_2}^n+\gamma_{3}{\lambda_3}^n,
   \end{equation}
where constants $\gamma_{1,2,3}$ are
\begin{equation}
\begin{array}{l}
  \gamma_1=C(0)
    \displaystyle
    \frac{\Gamma_2+\lambda_2 \lambda_3-\Gamma_1(\lambda_2+\lambda_3)}
    {(\lambda_1-\lambda_2)(\lambda_1-\lambda_3)},
  \\[6mm]
  \gamma_2=C(0)
    \displaystyle
    \frac{\Gamma_2+\lambda_1 \lambda_3-\Gamma_1(\lambda_1+\lambda_3)}
    {(\lambda_1-\lambda_2)(\lambda_2-\lambda_3)},
    \\[6mm]
  \gamma_3=C(0)
    \displaystyle
    \frac{\Gamma_2+\lambda_1 \lambda_2-\Gamma_1(\lambda_1+\lambda_2)}
    {(\lambda_1-\lambda_3)(\lambda_2-\lambda_3)},
\end{array}
\end{equation}
and $\lambda_{1,2,3}$ are roots of the characteristic equation of
the 3rd degree.  They can be either three real roots, or one real
and two complex conjugates. The solutions of the cubic equation are
quite lengthy and known and we will not write down them here. The
dependence of the correlation functions on the distance
between the elements of the chain is similar to those shown in Fig.
\ref{fig:2}.

\subsection{Examples of correlations in the $N$th
order chains }
\begin{figure}
\begin{center}
  \includegraphics[width=100mm]{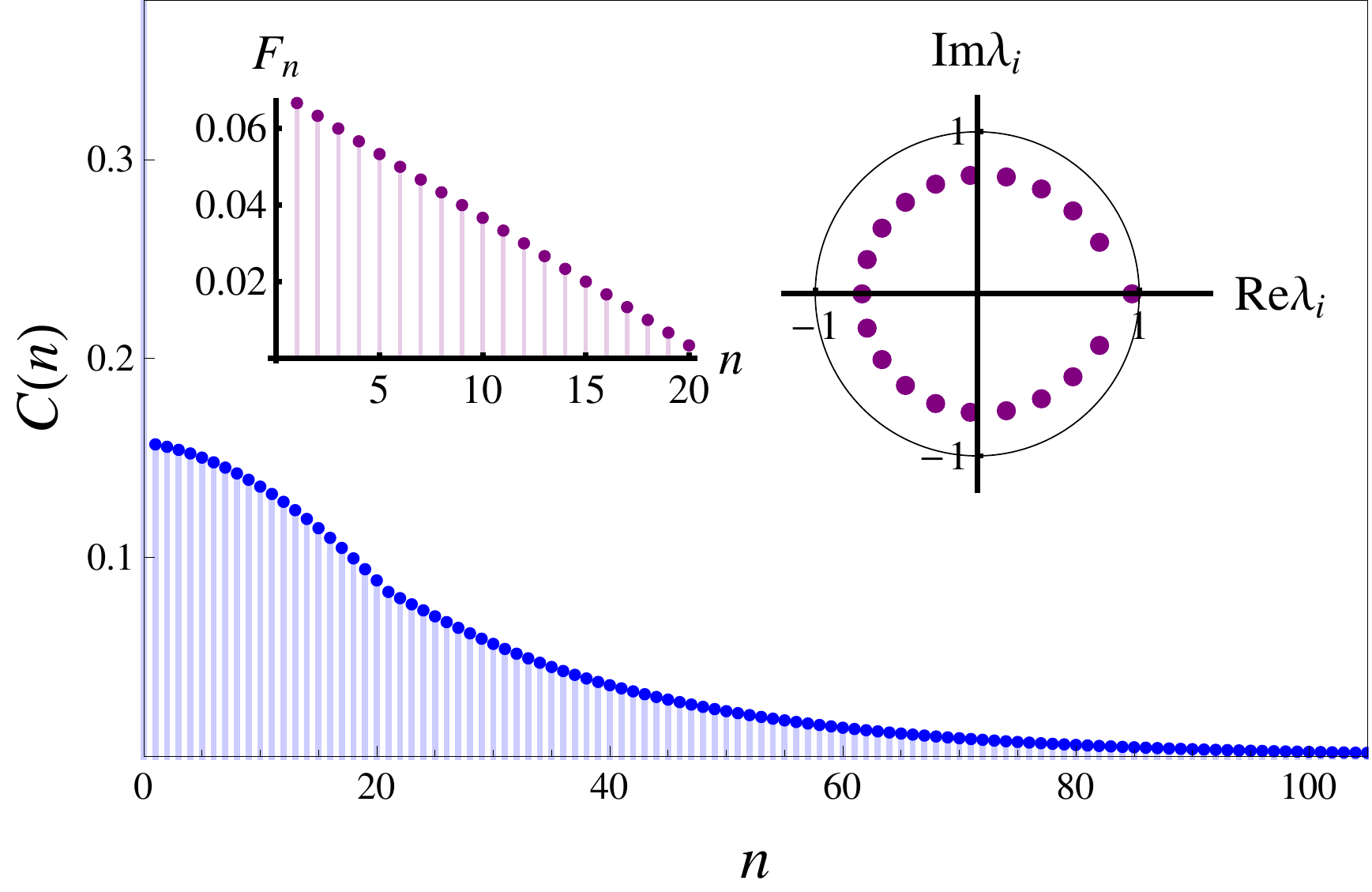}\hspace{5mm}
\end{center}
\hspace{-7mm}
  \caption{Linear memory function $F_n$ (top left),
    characteristic equation roots $\lambda$ on the complex plane (top right)
    and correlation function $C(n)$ (bottom).}
  \label{fig:5}
\end{figure}

\begin{figure}
\begin{center}
\hspace{-7mm}
  \includegraphics[width=100mm]{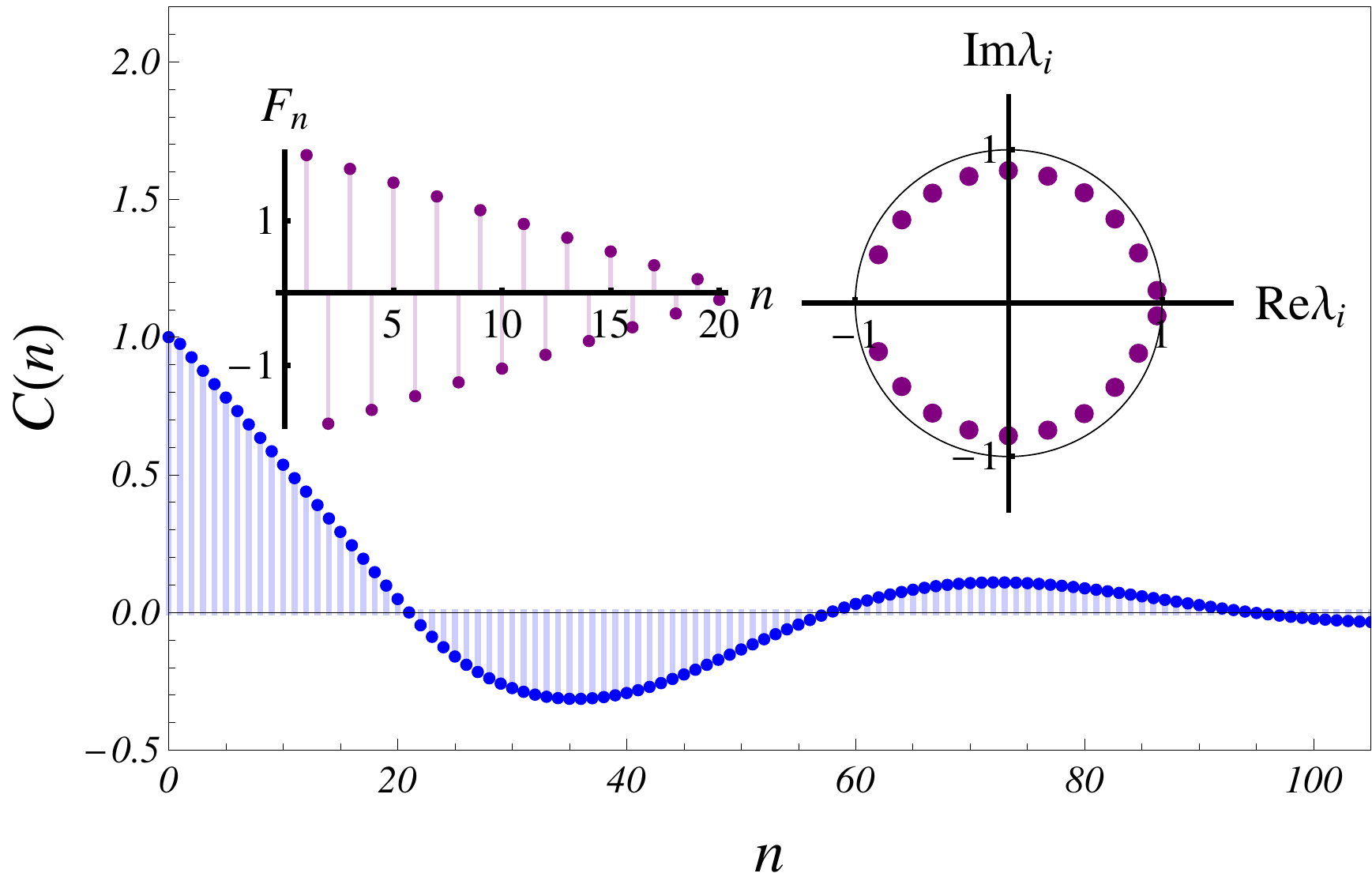}
\end{center}
  \caption{Alternating sign linear memory function $F_n$ (top left),
    characteristic equation roots $\lambda$ on the complex plane (top right)
    and correlation function $C(n)$ (bottom).}
  \label{fig:6}
\end{figure}

\begin{figure}
\begin{center}
\hspace{-7mm}
  \includegraphics[width=100mm]{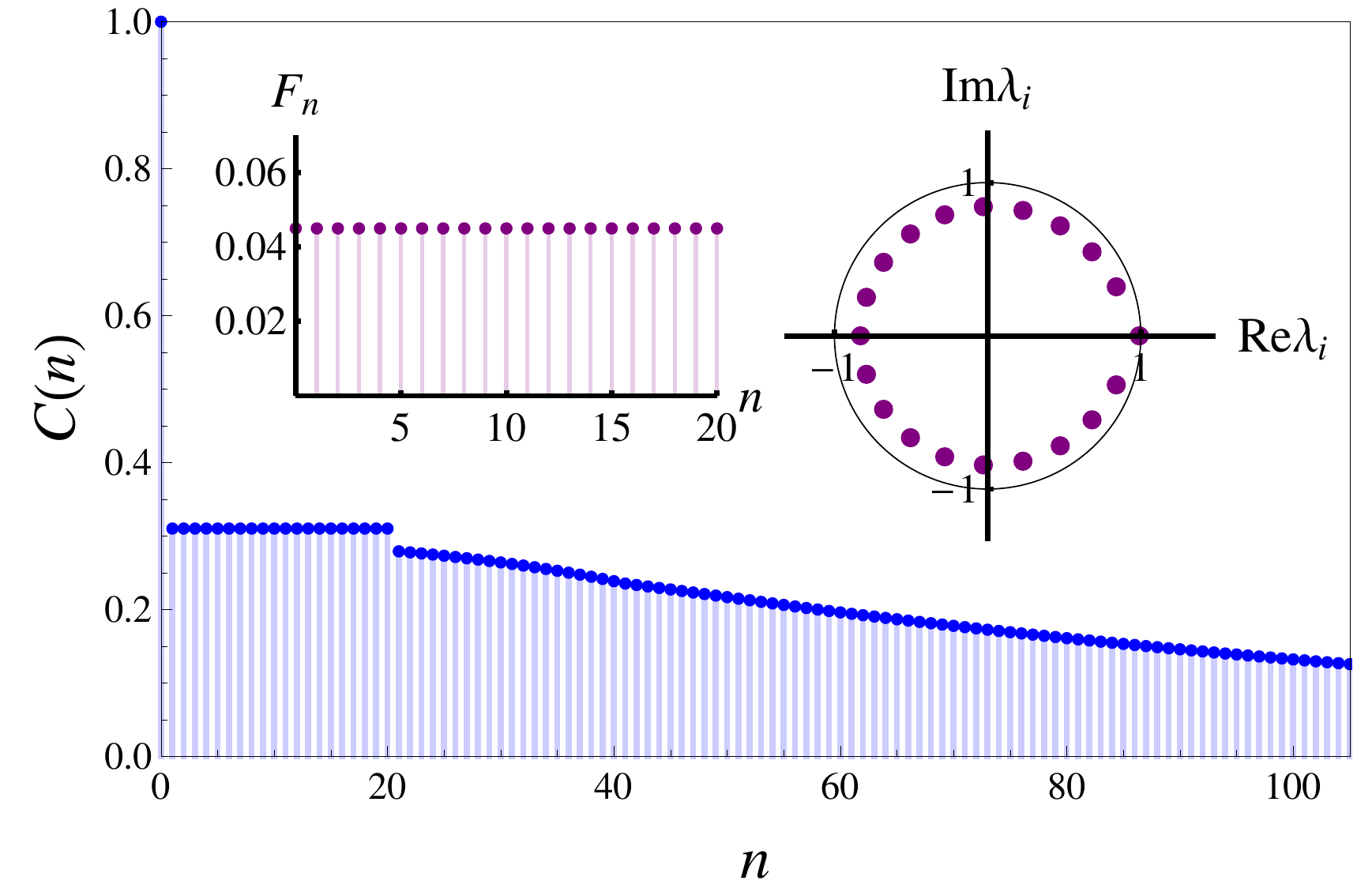}
\end{center}
  \caption{Step-wise  memory functions $F_n$ (top left),
    characteristic equation roots $\lambda$ on the complex plane (top right)
    and correlation function $C(n)$ (bottom).}
  \label{fig:7}
\end{figure}

\begin{figure}
\begin{center}
\hspace{-7mm}
  \includegraphics[width=100mm]{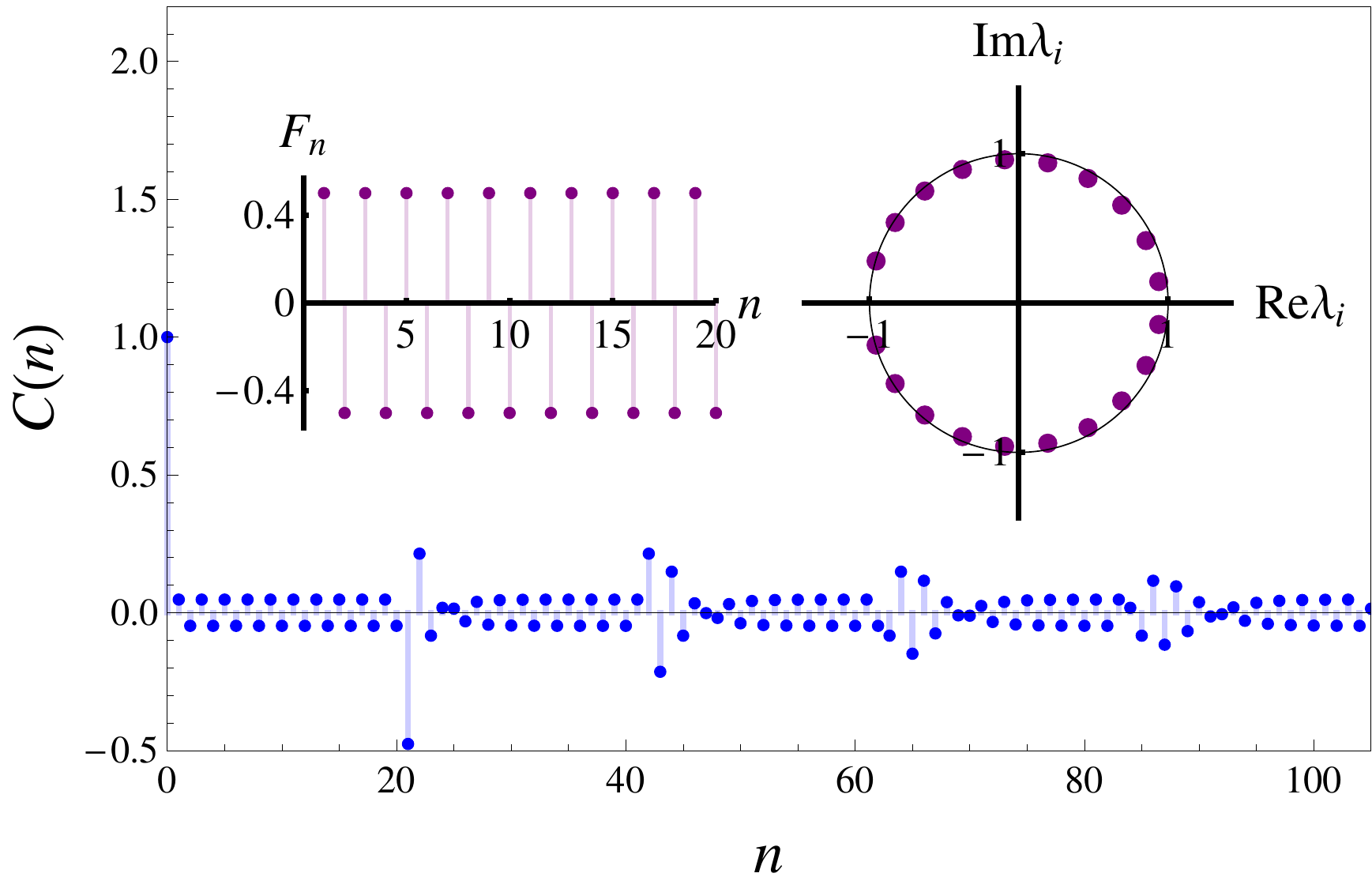}
\end{center}
  \caption{Alternating sign step-wise memory functions $F_n$ (top left),
    characteristic equation roots $\lambda$ on the complex plane (top right)
    and correlation function $C(n)$ (bottom).}
  \label{fig:8}
\end{figure}

Below, to illustrate possible correlation scenarios in the additive
Markov chains, we present plots of correlation functions for the
additive linear Markov chains of order $N=20$ for different types of
memory functions. In each of the Figs. \ref{fig:5}--\ref{fig:8},
the memory function is
shown at the top left, the roots of the polynomial equations
\eref{CharEq}, corresponding to the respective memory functions, are
at the top right and the dependence of the correlation function
\eref{GenSol} on the distance between the elements of the chain is
at the bottom.

\section{Numerical generation of the chain with a given 
correlation function \label{Sec:Generation}}

The main purpose of the developed theory is to elaborate a
reliable tool for the generation of numerical random sequences with
prescribed correlation characteristics.
The previous consideration concerned mostly the so-called 
\emph{direct} problem, that is, the problem of finding the correlation function
of the generated 
sequence with a given CPDF. In this section we address the
\emph{inverse} problem, namely, the problem of retrieving the CPDF of the sequence 
provided the correlation functions are given. For the linear 
high-order Markov chains \eref{def:lin0}, the inverse 
problem can be formulated as follows:

\textbf{Proposition 5}. To find the weight functions $f_{\mu}(x)$ we
should solve the system of equations \eref{eq:corrC1} and \eref{def:F}.

Up to here our theory was based on the ensemble 
statistical averaging. Now we would like to
change our method of reasoning and show a generation of \emph{one}
chain of random variables possessing the same statistical
characteristic as the an ensemble of chains. It is evident that the
generated sequence have to be \emph{ergodic}.

According to the ergodic theorem (see, e.g.,
Ref.~\cite{Shields,shir}), the finiteness of $N$ together with the
strict inequalities,
\begin{equation}\label{ergo}
 0 < \mathbb{P}\left( X_{n}=x_{n} |\left\{ X_{i}=x_{i}\right\}_{n-N}^{n-1}
  \right) < 1, \, \, n \in \mathbb{Z},
\end{equation}
provides ergodicity of the random sequence. The stationarity and
ergodicity are the sufficient conditions for formulation of the two
well known important statements of information theory: asymptotic
equipartition property (the Shannon-McMillan-Breiman theorem),
\cite{Cover} and the Kac's lemma \cite{Kac,Cec}. Analogously, we
should impose these conditions for generating the random stationary
ergodic sequence.

Let us demonstrate this with an example of numerical generation of
the Markov sequence with memory length $N=2$.

We choose the probability distribution function of the uncorrelated
chain in a symmetric ``triangular'' form, see Fig. 4 (left):

\begin{figure}
\begin{center}
  \includegraphics[width=120mm]{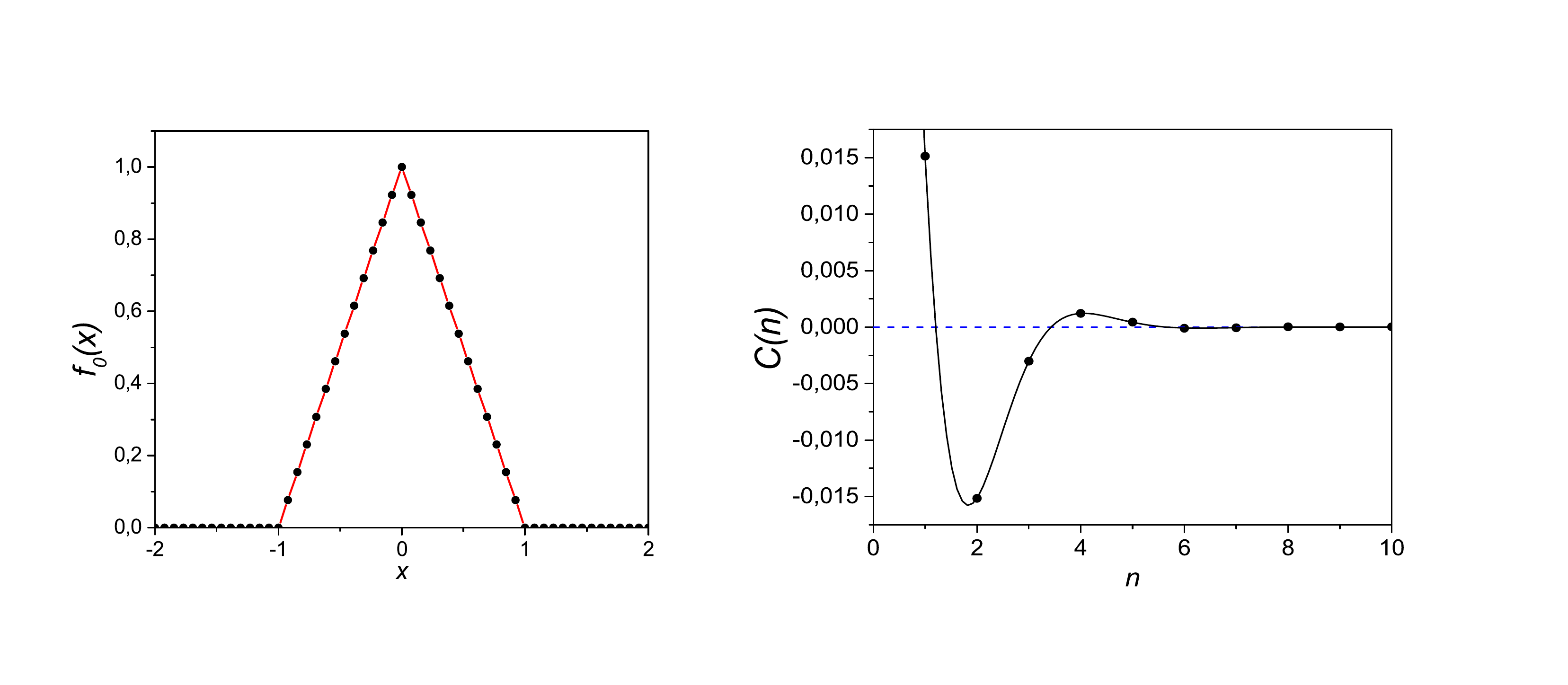}
\end{center}
  \caption{Numerical generation of the 2nd order chain:
    the one-point probability distribution function $f_0(x)$ (on the left);
    comparison of the numerical (points) and analytical (line) correlation functions (on the right).}
  \label{fig:4}
\end{figure}

\begin{equation}
  f_0(x) =
      \left\{
  \begin{array}{cr}
    1-|x|,& |x|<1,
  \\[3mm]
    0, & |x|\geqslant 1.
  \end{array}
  \right.
\end{equation}
This probability distribution function of uncorrelated chain
corresponds to the zero mean value of random variable $\langle X
\rangle=0$  and coincides with the one-point probability function
$P(X=x)$. The standard deviation is $C(0)=\sigma^2=1/6$. The values
of the memory function are chosen as $F_1=0,1$ and $F_2=-0,1$.

To generate, it remains to specify weight functions  $f_1$ and
$f_2$, which are part of the CPDF \eref{def:lin0}. As can be seen
from Eq.~ \eref{def:F}, there is a great freedom in the choice of
the functions $f_1$  and $f_2$  which ensure the given values of
$F_1$ and $F_2$. However, their form should be such that
\eref{def:lin0} yields a positive result for any value $x\in (-1;1)$
and  any combination of two preceding values $x_{1,2}\in (-1;1)$ .
This implies, in particular, that for the values of $x$  which give
close to zero values of  $f_0(x)$ (in our case, these values of $x$
are close to $\pm 1$) the absolute value of $f_{1,2}(x)$ should also
be close to zero. The simplest way to ensure this condition is to
choose $f_{1,2}(x)$  proportional to $f_{0}(x)$, but odd, to ensure
normalization conditions \eref{sum:f}:

\begin{equation}
  f_{1,2}(x) = a_{1,2}
      \left\{
  \begin{array}{cr}
    f_{0}(x),& x\geqslant 0,
  \\[2mm]
    -f_{0}(x), & x < 0,
  \end{array}
  \right.
\end{equation}
and the coefficient $a_{1,2}$ is then determined from the condition
\eref{def:F} to ensure the required $F_1$ and $F_2$. In the case
under consideration the triangular one-point distribution function
leads to $a_{1,2}=3F_{1,2}$.

Substituting now the required values of $F_1$ and $F_2$, we obtain
the CPDF \eref{def:lin0} and with its use generate the numerical
sequence. Having it constructed, we numerically calculate the
correlation functions. In Fig. \ref{fig:4} (right)  the points show
obtained values of $C(n)$. The solid line corresponds to the
analytical results for the correlation functions, obtained in
previous subsection. It is seen that the results of numerical
modeling of the correlated sequence perfectly coincide with those
obtained analytically.

\section{Conclusion}
We have shown that the additive linear CPDF given by
Eq.~\eref{def:lin0} can generate a stationary ergodic random
sequence with the one-point distribution function, Eq.~\eref{onePP},
and the pair correlation functions satisfying Eq.~\eref{eq:corrC}
with the boundary conditions Eq.~\eref{eq:corrC1}. The general
solution of Eq.~\eref{eq:corrC} is given by the linear combination
of the corresponding particular solutions Eq.~\eref{GenSol} with the
coefficients explicitly determined by Eq.~\eref{gamma}. The obtained
analytical solutions for the equations connecting the memory and
correlation functions are compared with the results of numerical
simulation. Examples of possible correlation scenarios in the
high-order additive linear chains are given.


%
\end{document}